\documentclass[a4paper,preprint]{jhep3}

\usepackage{amsmath}
\usepackage{bm}
\usepackage[dvips]{graphicx}

\newcommand{\be}{\begin{equation}}

\newcommand{\ee}{\end{equation}}
\newcommand{\bea}{\begin{eqnarray}}
\newcommand{\eea}{\end{eqnarray}}

\newcommand{\ha}{\frac{1}{2}}

\title{Self-consistent evaluation of quark masses in
three flavor crystalline color superconductivity}

\author{N.~D.~Ippolito, G.~Nardulli and M.~Ruggieri \\
Dipartimento di Fisica, Universit\`a di Bari, I-70126 Bari, Italia
\\and \\
I.N.F.N., Sezione di Bari, I-70126 Bari, Italia \\
E-mail: \email{nicola.ippolito@ba.infn.it}~,
\email{giuseppe.nardulli@ba.infn.it}~,
\email{marco.ruggieri@ba.infn.it}}

\preprint{BARI-TH 559/07}

\abstract{We present a self-consistent evaluation of the constituent
quark masses in the three flavor Larkin-Ovchinnikov-Fulde-Ferrell
(LOFF) phases of QCD, employing a Nambu-Jona Lasinio model. This
result allows to determine the window for values of the baryonic
chemical potential where the LOFF state is energetically favored.}


\begin{document}

\section{Introduction} The
phase diagram of three flavor QCD in the low temperature and high
density regime has recently received considerable theoretical
attention. Quark matter in these conditions is expected to be a
color superconductor
\cite{Collins:1974ky,Rajagopal:2000wf,Alford:1998mk}
 and this state of hadronic matter definitely deserves a
careful study. On the other hand the core of the compact stars is
the place where this exotic state of matter might be found.
Therefore the study of color superconductivity is interesting not
only theoretically, for it allows a deeper knowledge of the phase
diagram of QCD, but also phenomenologically, since this phenomenon
might be revealed in compact stellar objects.

 At very high densities  the color superconductor is expected to be in
the color-flavor locked state (CFL)~\cite{Alford:1998mk},
characterized by a spin zero homogeneous condensate with nine gapped
fermion excitations. When density is decreased, the electrical and
color neutrality conditions together with  finite quark mass effects
produce a mismatch of the Fermi surfaces of the pairing fermions. In
this case the CFL state has to be replaced by some less symmetric
phase. Among them the gapless CFL (gCFL)~\cite{Alford:2003fq}, the
gapless 2SC (g2SC)~\cite{Shovkovy:2003uu} and the $S=1$ color
superconductors \cite{Schafer:2000tw} have been extensively studied.

Besides the aforementioned color superconductive phases one can also
consider inhomogeneous states, produced when the fermion pair has
nonvanishing total momentum. This possibility was first discussed in
a condensed matter physics context~\cite{LOFF2}; the corresponding
phase is known as the LOFF state.  In QCD the LOFF color
superconductive phase has been studied firstly in the two flavor
case~\cite{Alford:2000ze} and then extended to the more interesting
three flavor
case~\cite{Casalbuoni:2005zp,Mannarelli:2006fy,Rajagopal:2006ig,Casalbuoni:2006zs}.
LOFF states are interesting since allow for pairing in conditions of
highly mismatched Fermi surfaces. Moreover it has been shown that
they do not suffer of the chromo-magnetic instability
problem~\cite{Giannakis:2004pf,Giannakis:2005vw,Giannakis:2005sa,Gorbar:2005tx,Ciminale:2006sm},
contrarily to the homogeneous gapless phases~\cite{Huang:2004bg}.

In the last years the studies on color superconductivity have been
improved by a self-consistent treatment of the constituent quark
masses in homogeneous phases
\cite{Steiner:2002gx,Buballa:2001gj,Abuki:2004zk,Ruster:2005jc,Blaschke:2005uj}.
These studies are based on phenomenological models, because their
aim is to study quark matter in an intermediate density regime where
the use of perturbative QCD is questionable. The favorite approach
is based on the  Nambu-Jona Lasinio (NJL) model \cite{Nambu:1961tp}
(see \cite{Klevansky:1992qe,Hatsuda:1994pi,Buballa:2003qv} for
reviews). This model embodies the symmetries of QCD. Although it
suffers many limitations, mainly due to the lacking of gluons as
dynamical fields, it is believed to capture some basic physics of
the  strong interactions and to describe the qualitative features of
the QCD phase diagram.

The aim of this paper is to evaluate in a self-consistent way the
constituent quark masses in the three flavor LOFF phase of QCD.  Our
approach is similar to that employed in
Refs.~\cite{Abuki:2004zk,Ruster:2005jc,Blaschke:2005uj} for the
homogeneous phases and therefore represents an extension of these
works to the LOFF color superconductive phases. To test the
robustness of our results we will consider two ways to treat the
ultraviolet cutoff at $\mu\neq 0$. In the first case we will use, as
usual in the literature, the same cutoff employed at $\mu=0$; in the
second case we treat the cutoff by the method discussed in
\cite{Casalbuoni:2003cs}, where a link between the cutoff and the
coupling constants is assumed. The two methods provide rather
similar results and in both cases windows of values of the chemical
potential are found where the LOFF phase is energetically favored.

 The plan of the paper is as follows. In
 Section \ref{2} we review the NJL model at $\mu=0$ and in
 particular
 we extend some of the results that were obtained in
 \cite{Casalbuoni:2003cs} for massless
 quarks to the case of massive quarks.
  In Section \ref{3} we extend these results at $\mu\neq 0$.
 We minimize the  free energy for the
 order parameters in the $\bar q q$ and  $ q q$ channels.
 For the $qq$ condensate we consider two different LOFF states: the
 one plane wave ansatz and two
structures based on cubic symmetry, one having 8 plane waves and
another having 16 plane waves. These two cases were considered in
Refs.
 \cite{Casalbuoni:2005zp,Mannarelli:2006fy} and
 \cite{Rajagopal:2006ig} respectively, assuming zero masses for the
 up and down quarks and taking the strange quark mass $M_s$ as a free parameter.
 The result of these papers is that the LOFF phase is favored for
 certain values of the parameters $M_s^2/\mu$. Since now we compute
 $M_s$ as a function of $\mu$ we are able to determine a range of
 values of $\mu$ where the LOFF phase is energetically favored.
  Finally in Section \ref{4} we draw our
conclusions.

\section{Nambu-Jona Lasinio  model at $\mu = 0$\label{2}}
In this Section we review the NJL model at zero quark chemical
potential. We consider a system of $u$, $d$ and $s$ quarks described
by the lagrangian
\begin{equation}
{\cal L} = \sum_f\bar\psi_f\,i\partial_\mu \gamma^\mu\psi_f + {\cal
L}_{mass}~+~{\cal L}_4~+~{\cal L}_6~;\label{eq:lagr1}
\end{equation}where the sum is over the flavors $f$ $(=1,2,3$ for $u,d,s$).
The mass term in the lagrangian is \be{\cal L}_{mass} ~=~-\,
\sum_fm_f\bar\psi_f\psi_f \ee and $m_f$ is the current mass. In this
paper we shall assume from the very beginning $m_u = m_d$. The NJL
four-fermion and six-fermion interaction Lagrangians
are~\cite{Klevansky:1992qe,Hatsuda:1994pi} \bea {\cal L}_4 &=&
G\sum_{a=0}^8\left[\left(\bar\psi \lambda_a \psi\right)^2 +
\left(i\bar\psi \gamma_5\lambda_a \psi\right)^2
\right]\label{eq:full4}~,\\ {\cal L}_6 &=& -K\left[{\rm
det}\bar\psi_f(1+ \gamma_5)\psi_{f'} \,+\,{\rm det}\bar\psi_f(1-
\gamma_5)\psi_{f'} \right] ~,\label{eq:full6} \eea where $\lambda_a$
are the Gell-Mann matrices in flavor space ($\lambda_0 =
\sqrt{2/3}~{\bm 1}_f$) and the determinant is in flavor space as
well.   ${\cal L}_4$  and ${\cal L}_6$ describe two-body and
three-body interactions respectively. In the Hartree approximation
one has for ${\cal L}_4$ the result
\begin{equation}
{\cal L}_4~=~ 4G\sum_{f}\sigma_f\,\bar \psi_{f} \psi_{f}~ -~
2G\sum_f \sigma_f^2~, \label{eq:MF1}
\end{equation}and for ${\cal L}_6$\be {\cal L}_6 ~=~-2\,K\,\sum_{f}
\sigma_{f+1}\sigma_{f+2}\,\bar \psi_{f}
\psi_{f}~+~4\,K\,\sigma_u\sigma_d\sigma_s~.\label{eq:MF1bis}\ee
 In deriving
these equations one assumes condensation only in the quark-antiquark
channel and treats consistently the number of colors $N_c$ assuming
that $KN_c^2\sim O(1)$, neglecting $1/N_c$ corrections
\cite{Klevansky:1992qe,Hatsuda:1994pi}. For each flavor $f$ one has
\begin{equation}
\sigma_f =  -i N_c\, \text{tr}S_f ~,
\end{equation}where $S_f$ is the propagator of the quark of flavour $f$,
$N_c$ is the number of colors, and the trace is on spinor indices
only. By definition $\sigma_4=\sigma_u,\,\sigma_5=\sigma_d$.

The six-fermion t'Hooft term mixes flavors and originates from the
$U_A(1)$ breaking contribution. At zero density it has the effect to
lift the degeneracy of the $\eta$ and $\eta\prime$ mesons. Note that
in a different approximation scheme it also modifies the
four-fermion giving rise to an effective four-fermion coupling
constant.  Using \eqref{eq:MF1} and \eqref{eq:MF1bis} the
self-consistent equations for the constituent quark masses at zero
baryon density read~\cite{Klevansky:1992qe,Ruster:2005jc}
\begin{equation}
M_f = m_f - 4 G \sigma_f~+2\,\,K\,\sigma_{f+1}\,\sigma_{f+2}\,,
\label{eq:masseMU0}
\end{equation}where
\be
\sigma_f=-\frac{3M_f}{\pi^2}\int_0^\Lambda\frac{p^2}{\sqrt{p^2+M_f^2}}\,dp~.
\ee

To fix the parameters of the model we use as an input the pion decay
constant $f_\pi=93$ MeV, the pion and kaon masses $m_\pi=135$ MeV
and $m_K=497$ MeV, the ratio \be\frac{<\bar s s
>_0}{<\bar u u>_0}\simeq 0.80\label{zerotto}\ee of the strange to the up chiral condensate,
and the values of the current quark masses. In detail the equation
for $f_\pi$  is\be f_\pi^2= -4i N_c M_u^2
I(0,M_u,M_u)~,\label{eq:fpi}\ee where \be I(k^2,M_1,M_2)=
\int\!\frac{d^4p}{(2\pi)^4}\frac{1}{[(p+k/2)^2-M_1^2][(p-k/2)^2-M_2^2]}~,
\label{eq:Iup}\ee From this relation we get $M_u(\Lambda)$. We note
that the quark condensates are given by~\cite{Hatsuda:1994pi}\be
\langle\bar \psi_f
\psi_f\rangle_0=\sigma_f~+~\frac{3m_f}{\pi^2}\int_0^\Lambda\frac{p^2}{\sqrt{p^2+m_f^2}}\,dp~.
\label{cond}\ee Using $m_u=m_d=5.5$ MeV,  and $M_u(\Lambda)$ we get
$\langle\bar \psi_f \psi_f\rangle_0$ as a function of the cutoff
$\Lambda$. To obtain $M_s(\Lambda)$ we use Eq. \eqref{cond}
evaluated for $f=3$ and Eq. \eqref{zerotto}. For the current strange
mass we take $m_s=135.4$ MeV, which is in the  range of values
allowed by QCD sum rules \cite{Colangelo:1997uy} and other methods
\cite{Yao:2006px} for the running current mass at the subtraction
point of 1 GeV and is to be preferred in view of a fit to the $\eta$
and $\eta^\prime$ mass, see below. The two constituent masses are
reported in Fig. \ref{Fig:masses} as a function of the cutoff
$\Lambda$.
\begin{figure}[ht!]
\begin{center}
\includegraphics[width=10cm]{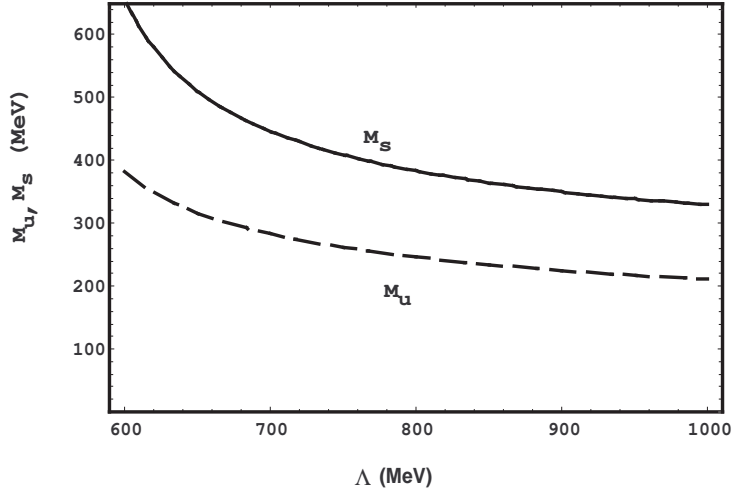}
\caption{\label{Fig:masses}Constituent quark masses $M_u$ and $M_s$
as a function of the cutoff $\Lambda$.}\end{center}
\end{figure}

\begin{figure}[ht!]\begin{center}
\includegraphics[width=10cm]{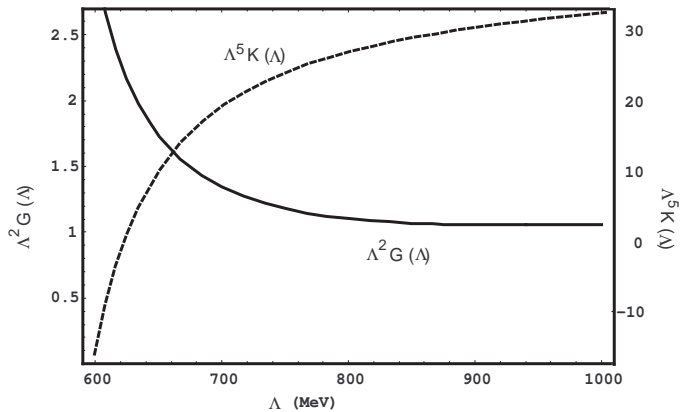}
\caption{\label{Fig:coupling} Couplings $\Lambda^2 G(\Lambda)$ and
$\Lambda^5 K(\Lambda)$ as a function of $\Lambda$.}
\end{center}\end{figure}

The equations giving the neutral pion and the neutral kaon masses
represent the mass-shell condition and are obtained in the random
phase approximation \cite{Klevansky:1992qe}. They are as follows:
\bea 1-\,(2\,G\,-\,K\sigma_s )\,\Pi(m_{\pi^0}^2,M_u,M_u) &= &0~,\label{eq:DispLawPi0}\\
1-\,(2\,G\,-\,K\sigma_u )\,\Pi(m_K^2,M_u,M_s) &= &0~,
\label{eq:DispLawK0} \eea with
 \be
\Pi(k^2,M_1,M_2)=-6i\int\frac{d^4p}{(2\pi)^4}\frac{k^2-4p^2+4M_1M_2}{[(p+k/2)^2-M_1^2][(
p-k/2)^2-M_2^2]} ~,\ee
 and allow to get $G$ and $K$ as a function of $\Lambda$.
 These results are in Fig. \ref{Fig:coupling}.

Finally we fix the value of the cutoff by requiring that the gap
equation Eq.~\eqref{eq:masseMU0} be satisfied for the strange quark:
on the l.h.s of Eq.~\eqref{eq:masseMU0} we use the $M_s(\Lambda)$
shown in Fig.~\ref{Fig:masses}; the same function is used on the
r.h.s. to evaluate $\sigma_s(\Lambda)$, and $G$ and $K$ are replaced
by the functions $G(\Lambda)$, $K(\Lambda)$ plotted in
Fig.~\ref{Fig:coupling}. If we denote by $\Lambda_0$ such a value of
the cutoff, valid at zero density and zero temperature, we get
\be\Lambda_0\simeq 643~~\rm {MeV}~.\label{cuto}\ee We note that by
this method one has another solution, with a smaller value of
$\Lambda_0$. However using equations analogous to
\eqref{eq:DispLawPi0}, \eqref{eq:DispLawK0} for the $\eta$ and
$\eta^\prime$ mesons (for the relevant formulae see Section IV B of
Ref. \cite{Klevansky:1992qe}) and the value \eqref{cuto} one gets as
a result $m_\eta=481$ MeV (exp. 548 MeV) and $m_{\eta^\prime}=924$
MeV (exp. 956 MeV), while the use of the smaller cutoff would
produce a worse fit to the experimental masses.
 As a check of the procedure we compute
also $f_K$ \cite{Klevansky:1992qe}. We get $f_K\simeq 98$ MeV (exp.
$112$ MeV). The values of the other parameters, computed at
$\Lambda_0$, are $M_u=M_d=324$ MeV, $M_s=521$ MeV and  $\langle\bar
uu\rangle_0=(-247~{\rm MeV})^3$. Finally we find \be G(\Lambda_0)
=\frac{ 1.81}{\Lambda_0^2}~,~~~~K(\Lambda_0) =
\frac{8.80}{\Lambda_0^5}~.\label{g0}\ee

\section{Constituent masses in the LOFF phases\label{3}} In this section
we  evaluate constituent quark masses and gaps for the color
superconductive LOFF phase of QCD. At finite chemical potential and
in presence of color condensation the lagrangian becomes
\begin{equation}
{\cal L} = \bar\psi\left(i\partial_\mu \gamma^\mu +
\hat\mu\gamma_0\right)\psi + {\cal L}_{mass} + {\cal L}_4 + + {\cal
L}_6+{\cal L}_\Delta~,\label{eq:lagr1MU}
\end{equation}
where ${\cal L}_{mass}$, ${\cal L}_4$ and ${\cal L}_6$ have been
discussed in Section \ref{2} and we have added a term ${\cal
L}_\Delta$ whose presence is responsible for color condensation. The
matrix $\hat\mu$ is a matrix diagonal both in flavor and color. It
depends on $\mu$ (the average quark chemical potential), $\mu_e$
(the electron chemical potential), and $\mu_3,\,\mu_8$ (color
chemical potentials) \cite{Alford:2003fq}. For color and electric
neutrality to be implemented it is sufficient to consider only these
chemical potentials, related as they are to the charge matrix and
the diagonal color operators $T_3 = \frac 1 2 {\rm diag}(1,-1,0)$
and $T_8 = \frac{1}{2 \sqrt 3 }{\rm diag}(1,1,-2)$ (in general one
should introduce a color chemical potential for each $SU(3)$ color
charge; however, as shown in~\cite{Buballa:2005bv}, for the
condensate with the color-flavor structure considered in this paper
it is enough to consider only $\mu_3$ and $\mu_8$, since the charges
related to the other color generators automatically vanish).
Therefore the matrix $\hat\mu$ is written as follows\be
{\hat\mu}_{ij}^{\alpha\beta} = \left(\mu \delta_{ij} - \mu_e
Q_{ij}\right)\delta^{\alpha\beta}+ \delta_{ij} \left(\mu_3
T_3^{\alpha\beta}+\frac{2}{\sqrt 3}\mu_8 T_8^{\alpha\beta}\right)
\label{9}\ee with $Q= {\rm diag} (2/3,-1/3,-1/3)$ ($i,j =1,3 $
flavor indices; $\alpha,\beta =1,3 $ colour indices).

The interaction term in the quark-quark channel ${\cal L}_\Delta$ is
\begin{equation}
{\cal L}_\Delta = \frac{3G}{4}\sum_{I=1}^3\left(\psi_{\alpha
i}^\dagger i\gamma_5\epsilon^{\alpha\beta I}\epsilon_{i j I} C
\psi_{\beta j}^*\right)\left(\psi_{\gamma m}^T C
i\gamma_5\epsilon^{\gamma\delta I} \epsilon_{mn I} \psi_{\delta n}
\right)~.\label{eq:LagrDelta1}
\end{equation}
Both ${\cal{L}}_\Delta$ and ${\cal{L}}_4$ in Eq.~\eqref{eq:full4}
can be obtained by a Fierz rearrangement of the four fermion
interaction with the quantum numbers of a single gluon exchange~(see
for example~\cite{Buballa:2003qv}). This procedure fixes the ratio
of the coupling constants, in the quark-antiquark and in the
quark-quark channel, to be equal to $3/4$. In principle an
additional contribution to ${\cal L}_\Delta$ arises from the
six-fermion interaction term, when treated in the mean field
approximation. Moreover, there exists also a repulsive interaction
in the symmetric color channel. In this paper we neglect both of
these terms for simplicity: the effects of these interactions are
expected to be small in the high density regime, to which we are
interested here.

In the mean field approximation one gets
\begin{equation}
{\cal L}_\Delta = -\frac{1}{2}\sum_{I=1}^3\left(\Delta_{I}({\bm
r})\psi_{\alpha i}^\dagger \gamma_5\epsilon^{\alpha\beta
I}\epsilon_{i j I} C \psi_{\beta j}^* + h.c. \right) +
\frac{\Delta_I({\bm r})\Delta_I^*({\bm
r})}{3G}~.\label{eq:LagrDelta2}
\end{equation}
In order to study the LOFF phase in a generic case we
assume
\begin{equation}
\Delta_{I}({\bm r}) = \Delta_I \sum_{m=1}^{P_I} \exp\left(2i{\bm
q}_m^I \cdot {\bm r}\right)~.\label{eq:ansatz1}
\end{equation}
and $P_I$ is the number of plane waves for each diquark condensation
channel and $2\,{\bm q}_m^I$ is the quark pair momentum. In the
sequel we consider only the case ${\bm q}_m^I = q_I \hat{\bm
n}_m^I$. Clearly $\Delta_1,~\Delta_2,~\Delta_3$ refer to $ds$,
$us$,$ud$ pairings respectively.

Once the lagrangian is fixed we compute the quark contribution to
the free energy in the Hartree approximation:
\begin{equation}
\Omega = \Omega_{e}~ +\Omega_n~ + \Omega_\Delta ~,\label{eq:grand1}
\end{equation}where $\Omega_{e}$ is the free energy of the ideal gas of
electrons: $\Omega_{e}=-\mu_e^4/12\pi^2$, and $\Omega_n$ is the
contribution of the unpaired phase:
\begin{equation}
\Omega_n= -4K\sigma_u\sigma_d\sigma_s+2G\!\sum_{f=u,d,s}\sigma_f^2
-\,2\,N_c\! \sum_{f=u,d,s}\int\!\frac{d^3p}{(2\pi)^3}\left\{E_f -
\left[E_f-\mu_f\right]\Theta\left(\mu_f-E_f\right)\right\}~,\label{eq:GrandNorm}
\end{equation}where $E_f = \sqrt{p^2 +
M_f^2}$. The constituent masses $M_f$ are here expressed in terms of
the $\sigma_f$ by means of \eqref{eq:masseMU0}.  We note that the
t'Hooft term gives rise to contributions that mix $qq$ and $\bar q
q$ condensates. Here, as in~\cite{Ruster:2005jc}, we neglect such
contributions for simplicity, since they are not expected to change
qualitatively the structure of the phase diagram. $\Omega_\Delta$ is
the contribution from the condensation in the diquark channel. We
write this last term in the Ginzburg Landau approximation, as this
is the scheme where the LOFF state of QCD with three flavors has
been studied so far, see \cite{Casalbuoni:2005zp},
\cite{Rajagopal:2006ig} and, for a test of the results of
\cite{Casalbuoni:2005zp}, also \cite{Mannarelli:2006fy}:
\begin{eqnarray}
\Omega_\Delta &= &\frac{2\mu^2}{\pi^2}\Biggl[\sum_I P_I \alpha_I
\, \Delta_I^*\Delta_I +\ha\Biggl(\sum_I
\beta_I(\Delta_I^*\Delta_I)^2
+\sum_{I>J} \beta_{IJ}\, \Delta_I^*\Delta_I\Delta_J^*\Delta_J\Biggr) \nonumber\\
&&+\frac{1}{3}\Biggl(\sum_I \gamma_I(\Delta_I^*\Delta_I)^3
+\sum_{I\neq J}\gamma_{IJJ}\,
\Delta_I^*\Delta_I\Delta_J^*\Delta_J\Delta_J^*\Delta_J
+\gamma_{123}\,\Delta_1^*\Delta_1\Delta_2^*\Delta_2\Delta_3^*\Delta_3\Biggr)\Biggr]
\label{eq:GLexpansion}~.
\end{eqnarray}We include terms up to sixth order in the gap
paramenters, consistently with the results of
\cite{Rajagopal:2006ig}.

It is useful to note that, since we are interested in the three
flavor paired quark matter, the chemical potential of the strange
quark has to be larger than its constituent mass, otherwise no Fermi
sphere for $s$ quarks would exist.  In order to determine the ground
state we have to minimize $\Omega$ with respect to $\sigma_f$,
$\Delta_I$ and $q_I$ under the conditions of electrical and color
neutrality. In the region of interest we expect $\mu_e \approx
M_s^2/4\mu$ and $\mu_3 \approx \mu_8 \approx 0$, i.e. the results
for the normal non-superconductive phase. Therefore we
 assume from the very beginning $\Delta_1 = 0$, $q_2 = q_3
\approx 0.6(\mu_d-\mu_u)$ and $\Delta_2 = \Delta_3
\equiv\Delta$~\cite{Casalbuoni:2005zp,Rajagopal:2006ig}. Moreover we
are interested to structures with $P_2 = P_3 \equiv P$. With these
approximations one can write~\cite{Rajagopal:2006ig}
\begin{equation}
\Omega_\Delta =
\frac{2\mu^2}{\pi^2}\left[2P\alpha(\delta\mu)\Delta^2 +
\frac{\bar\beta_{eff}}{2\delta\mu^2}\Delta^4 +
\frac{\bar\gamma_{eff}}{3\delta\mu^4}\Delta^6 \right]~,
\end{equation}
where $\bar\beta_{eff}$ and $\bar\gamma_{eff}$ are dimensionless
coefficients that do not depend on $\delta\mu = M_s^2/8\mu$ and are
evaluated in~\cite{Rajagopal:2006ig} for several crystal structures.
In the following we refer to the values reported in Table II
of~\cite{Rajagopal:2006ig}. We notice that the parameters
$\bar\beta_{eff}$, $\bar\gamma_{eff}$ do not depend on the
ultraviolet cutoff. Furthermore, the cutoff dependence of the
quadratic coefficient in $\Omega_\Delta$ is replaced by the
dependence on the BCS gap parameter, which we calculate for each
value of $\mu$; at the minimum one has
\begin{equation}
\alpha(\delta\mu) =
-\frac{1}{2}\text{log}\left(\frac{\Delta_{2SC}^2}{4\delta\mu^2(\eta^2-1)}\right)~;
\end{equation}
in the above equation $\eta = 1.1997$ and $\Delta_{2SC}$ denotes the
gap parameter in the 2SC phase; it is related to the gap of the CFL
phase with massless flavors $\Delta_0$ by the relation
\begin{equation}
\Delta_{2SC} = 2^{1/3}\Delta_0~.
\end{equation}
Keeping in mind these approximations we are left with the four
equations ($f=u,d,s$)
\begin{equation}
\frac{\partial\Omega}{\partial \sigma_f} = 0 ~, ~~~~~
\frac{\partial\Omega}{\partial \Delta} = 0~,
\label{eq:minimizOmega1}
\end{equation}
which have to be solved simultaneously.
Eqs.~\eqref{eq:minimizOmega1} are equivalent to the Schwinger-Dyson
equations for the proper self-energy of the quarks; in particular
for $\Delta=0$ one gets the equations for the constituent quark
masses in the normal Fermi liquid phase, at finite chemical
potential \cite{Asakawa:1989bq}. We note that we have adapted the
numerical values of the parameters of the Ginzburg Landau expansion
to the numerical values for $\mu$ used in the present paper.

\begin{figure}[ht!]
\begin{center}
\includegraphics[width=12cm]{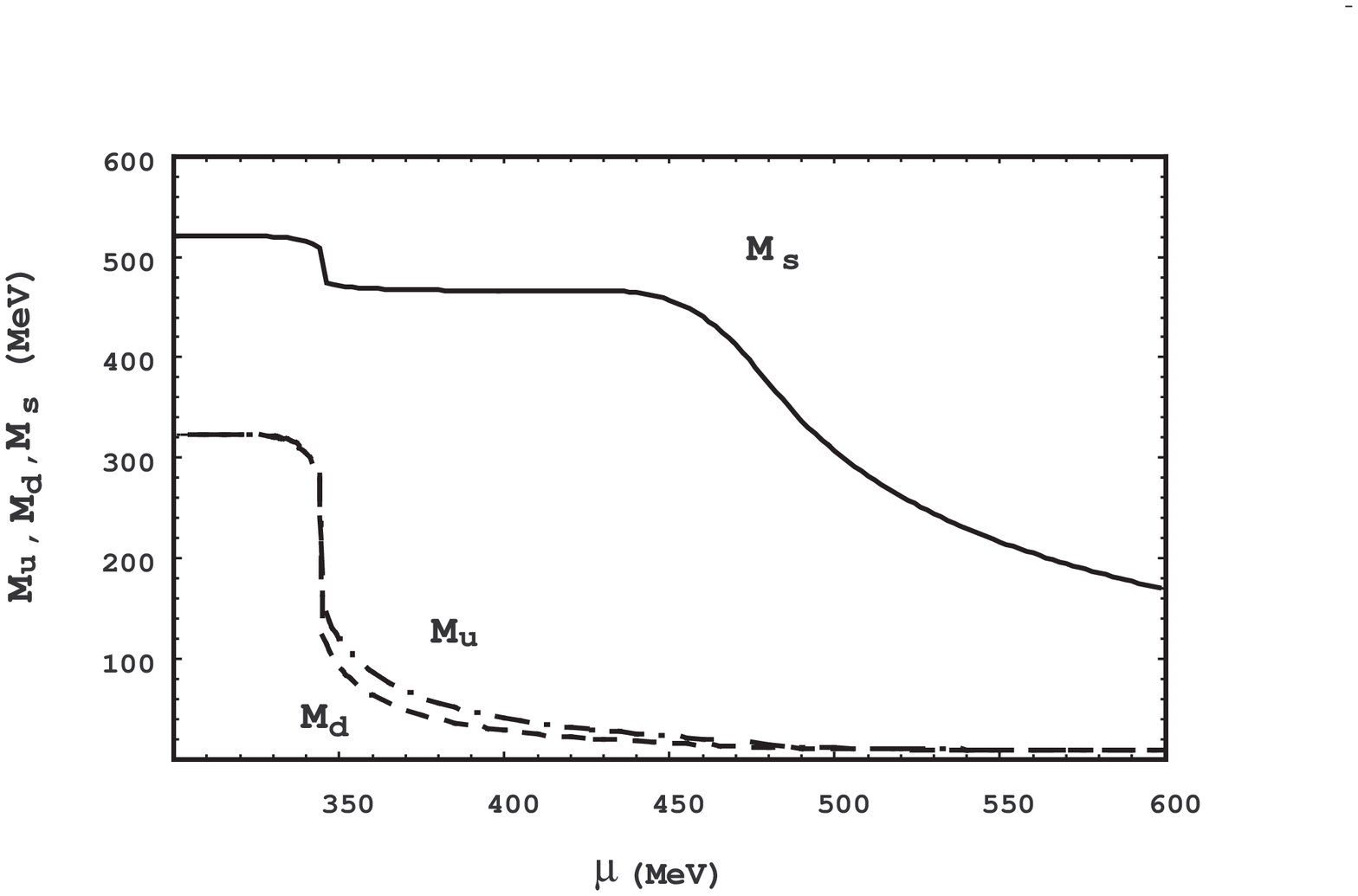}
\end{center}
\caption{\label{Fig:massaSmu} Constituent quark masses $M_s$,  $M_d$
and $M_u$  as functions of the baryon chemical potential $\mu$.}
\end{figure}

Numerically we find that $\Omega_\Delta/\Omega_n \approx 10^{-4}$
for the values of $\mu$ of interest. As a result the constituent
quark masses in the LOFF phases do not differ significantly from the
ones in the unpaired phase ($\Omega_\Delta$ gives a shift in the
masses at most of the $2$ percent from the normal phase result). The
results are shown in Fig.~\ref{Fig:massaSmu}, where we plot the
constituent quark masses versus the chemical potential $\mu$. Note
that in this figure we use for the ultraviolet cutoff the value
\eqref{cuto}. One can note a first order phase transition at
$\mu\approx 350$ characterized by a discontinuity in the values of
the constituent light quark masses, and a second order transition at
$\mu\approx500$ MeV (at $\mu\approx 350$ MeV there is a
discontinuity of the strange quark mass too: it is related to the
t'Hooft interaction term which links the strange quark mass to the
light quarks condensates, see Eq.~\eqref{eq:masseMU0}). As discussed
in \cite{Klevansky:1992qe}, the order of the phase transitions at
$T=0$ and large $\mu$ depends in NJL models with normal Fermi liquid
behavior on the approximations and the choice of the parameters.
This holds also in our case. The difference between $M_u$ and $M_d$
arises from the finite value of $\mu_e$ for $\mu>350$ MeV.  We note
that in the range of $\mu$ of interest for the LOFF phase ($\mu\sim
$ 500 MeV) the condensates $<\bar uu
>_0$ and $<\bar d d >_0$ vanish, which implies that the constituent
masses $M_u$ and $M_d$ are negligible in comparison with to $M_s$.
As a consequence the values of the gap parameters depend only on the
strange quark mass. As stressed in the introduction, differently
from the previous analyses
\cite{Casalbuoni:2005zp,Mannarelli:2006fy,Rajagopal:2006ig} here the
strange quark mass is not treated as a free parameter, but
determined self-consistently.

Using these results we are able to compute the interval
($\mu_1\,,\mu_2)$ of values of $\mu$ where the LOFF phase is
favored. To begin with we examine the simple case of one plane wave
per condensate (denoted as 2PW in~\cite{Rajagopal:2006ig})
corresponding to $P_2 = P_3 =1$ in Eq.~\eqref{eq:ansatz1}. In
Ref.~\cite{Casalbuoni:2005zp,Rajagopal:2006ig,Mannarelli:2006fy} it
was found that this LOFF state is energetically favored for the
following values of the ratio $M^2_s/\mu$,
\begin{equation}
4.80\Delta_0 \le \frac{M_s^2}{\mu} \le
7.56\Delta_0~,\label{eq:window1}
\end{equation}
where $\Delta_0$ is the CFL gap in the chiral limit. It is given by
\begin{equation}
\Delta_0 =
2^{2/3}(\Lambda_0-\mu)~\text{exp}\left(-\frac{\pi^2}{6G(\Lambda_0)\mu^2}\right)
\end{equation}
with $\Lambda_0$ and $G(\Lambda_0)$ given in \eqref{cuto} and
\eqref{g0}. We note that at the reference value of $\mu=500$ MeV we
find $\Delta_0 \approx 50$ MeV.

\begin{figure}[h!]\begin{center}
\includegraphics[width=10.5cm]{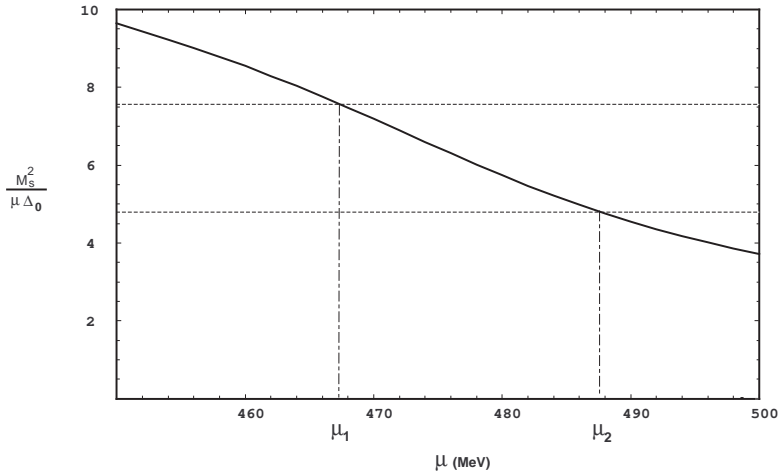}
\caption{\label{Fig:window1} Solid line:
$\displaystyle\frac{M_s^2}{\mu\Delta_0}$versus the baryon chemical
potential $\mu$ (MeV) for the LOFF phase with one-plane-wave
structure. The UV cutoff is given by Eq.~(2.17). The points
$\mu_1,\mu_2$ define the range of $\mu$ where the LOFF phase
prevails. For $\mu\,<\,\mu_1$ the favored phase is the normal one.
The upper limit is $\mu_2$ is obtained comparing the free energies
of the gCFL and the LOFF phase (see text).}\end{center}
\end{figure}

The LOFF window is defined by the two points where the horizontal
dotted lines cross the vertical axis in Fig.~\ref{Fig:window1}. The
curve represents
 $M_s^2/(\mu\Delta_0)$ as a function of $\mu$, with
$M_s$ and $\Delta_0$ both calculated self-consistently as functions
of the average quark chemical potential $\mu$. The dotted horizontal
lines intersect the curve at the values $\mu_1,\mu_2$ depicted in
the figure. In this way we obtain that the LOFF phase is favored in
comparison with the normal phase (this comparison fixes $\mu_1$) and
the gCFL phase (this latter comparison fixes $\mu_2$) for
$\mu\in(\mu_1,\,\mu_2)$, i.e. \be 467\,~\text{
MeV}\,<\,\mu\,<\,488~\text{ MeV} ~.\label{una}\ee Therefore there
exists a small but finite window in $\mu$ where the LOFF phase is
favored in comparison with the normal state and the unstable gCFL
phase.

As stressed in the introduction, the analysis of
\cite{Rajagopal:2006ig} shows that more complicated crystalline
structures can lead to a free energy smaller than the 2PW case.
In~\cite{Rajagopal:2006ig} it was found by a Ginzburg Landau
expansion  that a crystalline color superconductive phase exists in
the following interval:
\begin{equation}
2.88\Delta_0 \le \frac{M_s^2}{\mu} \le
10.36\Delta_0~.\label{eq:window2}
\end{equation}
In more detail, for $2.88\Delta_0 \le M_s^2/\mu   \le 6.20\Delta_0$
the ground state of three flavor quark matter is the CubeX. In this
structure $P_2 = P_3 = 4$; for each pairing channel the wave vectors
$\{{\bm q}_I\}$ form a square, and the two squares are arranged in
such a way that they point to the vertices of a cube. In the
remaining region the favored structure is the 2Cube45z in which $P_2
= P_3 = 8$; each wave vector set $\{{\bm q}_I\}$ forms a cube, and
the two cubes are rotated by 45 degrees around an axes perpendicular
to one of the faces of the cube. Using Eq.~\eqref{eq:window2} and
the results for the constituent strange quark mass, plotted in
Fig.~\ref{Fig:massaSmu}, we obtain the LOFF phase is favored for the
values of $\mu$ in the window $(\mu_1,\mu_2)$ defined by\be
442\,~\text{ MeV}\,<\,\mu\,<\,515~\text{ MeV} ~.\label{cubo}\ee  We
conclude that there exists  a window in $\mu$, larger than that of
the 2PW structure, where the cubic LOFF phase is energetically
favored in comparison with the normal state, the 2PW LOFF phase, and
the unstable gCFL phase.

\begin{figure}[h!]
\begin{center}
\includegraphics[width=10.5cm]{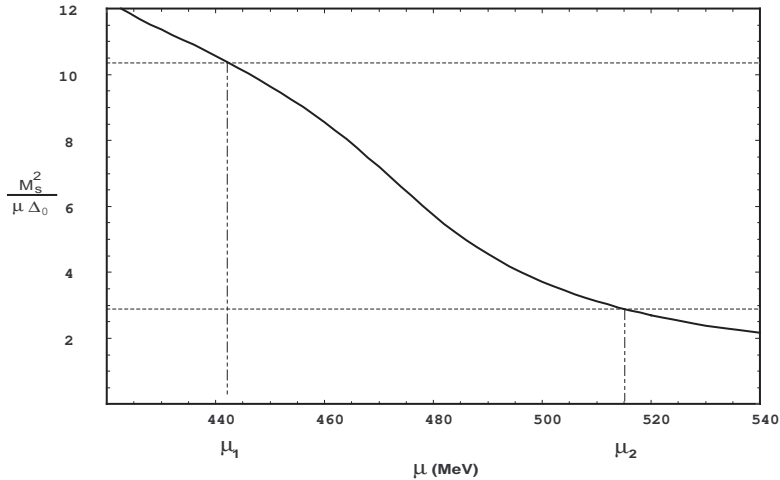}
\end{center}
\caption{\label{Fig:cubic} Solid line: ratio
$\displaystyle\frac{M_s^2}{\mu\Delta_0}$ versus the baryon chemical
potential $\mu$ (MeV) for the LOFF phase with cubic crystalline
structure. The UV cutoff, $\mu_1,\,\mu_2$ as in Fig.~4.}
\end{figure}

The results obtained here for the constituent quark masses are in
agreement with those obtained in Ref.~\cite{Ruster:2005jc}. As
already stressed, this is due to the fact that the contribution of
the quark-quark condensation to the free energy is negligible, and
the constituent quark masses are almost equals to their value in the
phase with $\Delta=0$.

Let us now discuss the dependence of these results on the cutoff.
Thus far we have used an ultraviolet cutoff
 independent of the quark chemical potential and numerically equal to
 the value of the model at $\mu=0$ given by eq. \eqref{cuto}. For large
 values of $\mu$ (ie $\mu\gg 500$ MeV) this choice
 leads to a  non-monotonic behavior of $\Delta$ with
 $\mu$. Therefore in \cite{Casalbuoni:2003cs} a different procedure
 was suggested. It is based on the following choice:
$\Lambda(\mu)=\mu+\delta$, with $\delta=c\mu$. The constant $c$ was
fixed imposing that the values of the gap are almost independent of
$c$. In this way it was found \be c=0.35\pm 0.10~,\label{c}\ee which
is what we assume here, though other choices are
possible~\cite{Baldo:2006bt}. Together with this choice of the
cutoff at $\mu\neq 0$ one has to use values for the couplings $G$
and $K$ computed at the same value of $\Lambda$ (the dependence can
be read from fig. \ref{Fig:coupling}). To test the robustness of our
computation we have therefore implemented this second choice of the
cutoff. The results we obtain are as follows. The values of the
masses are almost identical to the ones found with the cutoff
$\Lambda_0$, see Fig. \ref{Fig:massaSmu}. As to the windows where
the LOFF phase prevails, for $c=0.35$ and the case of the
one-plane-wave structure we find $ 468\,\text{
MeV}\,<\,\mu\,<\,481~\text{ MeV}$ instead of \eqref{una}; for the
case of the cubic structure we find $445\,\text{
MeV}\,<\,\mu\,<\,640~\text{ MeV} ~$ instead of the results
\eqref{cubo}. Varying $c$ in the window~\eqref{c} we obtain similar
results. Though numerically different from the results obtained with
the cutoff $\Lambda_0$, the results found with the alternative
cutoff procedure do not alter the overall conclusion, ie that there
exists a window of values of $\mu$ where the LOFF phase is
energetically favored. For the case of the single plane wave the
range is between $\approx 470$ and 480 MeV. For the case of the
cubic structure the window is larger, with a lower limit at
$\approx445$ MeV and an upper limit more model dependent, but, in
any case not smaller that 545 MeV.

Let us finally note that the crystalline color superconductive phase
does not have a free energy smaller than gCFL for all the values of
the strange quark mass. If the gCFL phase has smaller free energy,
as it is unstable, it must be replaced by a new state, most likely
to be a current-carrying meson condensate state and/or gluonic
phases~\cite{Huang:2005pv}.

\section{Conclusions\label{4}}
We have evaluated the three flavor QCD LOFF phase by a
self-consistent computation of the quark masses. We have adopted two
different methods to cutoff the theory in the ultraviolet regime,
with results that are basically consistent. We have determined the
LOFF window as a function of the baryon chemical potential only,
instead of the usual $M_s^2/\mu$ representation where the
constituent quark masses are taken as external parameters. We find
that  there exist ranges of values of the quark chemical potential
where the crystalline color superconductive phase has a free energy
smaller than the homogeneous and the normal phases.. We find the
range of $\mu$ given by Eq. \eqref{una} for the one plane wave
structure of the LOFF condensate, with the the cutoff \eqref{cuto},
and a slightly smaller range for the alternate cutoff $\Lambda =
\mu+c\mu$ with $c$ in \eqref{c}. As shown in \cite{Rajagopal:2006ig}
the cubic structure is energetically favored in comparison to the
single plane wave. The numerical results of \cite{Rajagopal:2006ig}
might be called into question since the phase transition to the
normal phase for the cubic structures is first order and therefore
the use of the Ginzburg Landau approximation is questionable in this
case. In any event also for the cubic structure there exists a range
of values of the quark chemical potential, expressed by Eq.
\eqref{cubo} and, as expected, larger than the range \eqref{una},
where the cubic crystalline LOFF structure is energetically favored
(for the alternate cutoff the range is even larger). Let us finally
note that, as shown by Fig. \ref{Fig:massaSmu}, in the region $350$
MeV $<\mu<445$ MeV the Fermi surfaces of the $u$ and $d$ quarks are
present, whereas the one of the strange quark is not. Even though
the LOFF state with three flavors is not possible in this case,
still there exists  the possibility of a LOFF state with a $\langle
ud \rangle$ color condensate. This case has been well studied in the
literature \cite{Alford:2000ze}. We have computed in this case the
difference in chemical potential between the $u$ and $d$ quarks:
$\delta\mu=\mu_e/2$, and we have noted that in correspondence of
these values one or more crystalline structures
 of the LOFF state with two flavors (and more than one plane wave)
 are possible. Therefore the window for $\mu$ available for the LOFF color
 superconductivity is actually larger than shown by figures \ref{Fig:window1} and
 \ref{Fig:cubic}.

Our results could be relevant for astrophysical applications of the
LOFF phase, because the ranges of $\mu$ we have found approximately
coincide with those considered for the core of compact stellar
objects. If the density in some region of a compact star do
correspond to baryon chemical potential belonging to the
range~\eqref{cubo}, one should conclude the quark matter is in an
inhomogeneous color superconductive state with a cubic crystal
structure. For a better understanding of the role of the QCD LOFF
phase in compact stars it is however of vital importance to compute
the transport properties of the crystalline three flavor color
superconductor. Since in the LOFF phase both the rotational and the
translational symmetries are broken by the condensate, phonons
appear in the spectrum. Recently the effective lagrangian of such
phonons has been evaluated in~\cite{Mannarelli:2007bs}. It has been
found that the crystalline color superconductor responds to shear
stress as a very rigid solid because of its high shear modulus; the
results of~\cite{Mannarelli:2007bs} show that some pulsar glitches
may originate within a quark matter core deep within a neutron star.
Beside the phonons, in the LOFF phase there exist an octet of
pseudo-Goldstone bosons related to the breaking of $SU(3)_A$: it
would be interesting to investigate if a kaon condensation occurs in
the ground state (in the CFL phase this problem has been extensively
studied, see for example~\cite{Bedaque:2001je}). We will come back
to this problem in the future. Besides the transport coefficients,
it might be interesting the evaluation of the cooling curves of a
compact star whose core is in a cubic LOFF state (the simple case of
two plane waves has been studied in~\cite{Anglani:2006br}). Another
interesting investigation is the use of different density dependent
cutoffs as in Ref.~\cite{Baldo:2006bt}. We leave also this study to
future work.

\acknowledgments We thank R.~Anglani, D.~Blaschke, M.~Buballa,
R.~Casalbuoni, M.~Ciminale, R.~Gatto, V.~Laporta, M.~Mannarelli,
K.~Rajagopal and R.~Sharma for enlightening discussions and
comments.

\end{document}